\documentclass[prb,twocolumn,showpacs,floatfix]{revtex4}
\usepackage{graphicx}

\def\be{\begin{equation}}
\def\ee{\end{equation}}
\def\bea{\begin{eqnarray}}
\def\eea{\end{eqnarray}}

\begin{document}

\title{Two-dimensional tunnel correlations with dissipation}

 \author{A. K. Aringazin}
 \email{aringazin@mail.kz}
 \altaffiliation[Also at ]
 {Kazakhstan Division, Moscow State University, Moscow 119899,
Russia.}
 \affiliation{Department of Theoretical Physics, Institute
for Basic Research, Eurasian National University, Astana 473021,
Kazakhstan}
 \homepage{http://www.emu.kz}

 \author{Yuri Dahnovsky}
 \email{yurid@uwyo.edu}
 \affiliation{Department of Physics and Astronomy, P.O. Box 3905,
University of Wyoming, Laramie, WY 82071, USA}

 \author{V. D. Krevchik}
 \email{physics@diamond.stup.ac.ru}
 \affiliation{Department of Physics,
Penza State University, 40 Krasnaya St., Penza 440017, Russia}

\author{A. A. Ovchinnikov}
 \affiliation{Joint Institute of Chemical Physics,
Kosygin Street 4, 117334 Moscow, Russia}


 \author{M. B. Semenov}
 \email{physics@diamond.stup.ac.ru}
 \altaffiliation[Also at ]{Institute for Basic Research, P.O. Box
1577, Palm Harbor, FL 34682, USA.}
 \affiliation{Department of
Physics, Penza State University, 40 Krasnaya St., Penza 440017,
Russia}

 \author{K. Yamamoto}
 \affiliation{Research Institute of International Medical Center of
Japan, Tokyo, Japan}

\date{August 20, 2003}

\begin{abstract}

Tunneling of two particles in synchronous and asynchronous regimes
is studied in the framework of dissipative quantum tunneling. The
critical temperature $T_c$ corresponding to a bifurcation of the
underbarrier trajectory is determined. The effect of a heat bath
local mode on the probability of two-dimensional tunneling
transfer is also investigated. At certain values of the
parameters, the degeneracy of antiparallel tunneling trajectories
is important. Thus, four, six, twelve, etc., pairs of the
trajectories should be taken into account ({\em a cascade of
bifurcations}). For the parallel particle tunneling the
bifurcation resembles phase transition of a first kind, while for
the antiparallel transfer it behaves as second order phase
transition. The proposed theory allows for the explanation of
experimental data on quantum fluctuations in two-proton tunneling
in porphyrins near the critical temperature.
\end{abstract}

\pacs{03.65 Xp, 03.65 Sq, 31.15 Gy, 31.15 Kb, 73.40 Gk, 82.20 Xr}

\maketitle

\section{\label{Sec1} Introduction}

Quantum tunneling dynamics of a particle interacting with a heat
bath is one of the important problems of modern condensed matter
physics.
\cite{1,3,4,5,6,7,8,9,10,11,12,13,14,15,16,17,18,19,20,21,22,23,24,25,26,27,28,29,Ankerhold,dos}
The interest to this problem is related to studies in
low-temperature superconductive tunnel junctions, \cite{10,11,14}
a dissipative quantum tunneling in crystals, \cite{18} and
low-temperature chemical reactions. \cite{1,3,4,5,23,24,dos,6} In
low dimensional systems an effective mass approximation often
fails. Thus, quantum tunneling with dissipation becomes an
important tool in the description of electron transfer.
\cite{8,18} In many physical processes tunneling of {\em two}
particles occurs. For example, Semenov and Dakhnovskii
\cite{24,dos} found the bistability of the tunneling trajectories
in a two-proton transfer. Later, Benderskii and co-workers
extended their investigations to different two-dimensional
potentials. \cite{6} Some features of the two-dimensional
tunneling dynamics were studied in interacting Josephson
junctions. \cite{11}

Another example of two-particle tunneling is a two-proton transfer
in porphyrin systems. \cite{23,dos,24} Porphyrins are important
molecules in Biology \cite{fr} and a new area of Electronics --
molecular wires and devices. \cite{tag} There are experimental
data \cite{3,4,5,Mamaev} that clearly indicate that a tunneling
instability occurs at some critical temperature. Such a behavior
in the rate constant demonstrates the existence of bifurcations in
two-dimensional underbarrier trajectories. Apparently, this effect
requires a thorough investigation when the protons interact with a
thermal bath.

In this paper, we continue to study  bifurcation effects in two
dimensional tunneling. In particular, we consider  two-proton
correlations of different types within the framework of
dissipative quantum tunneling (instanton) approach. Moreover, we
discuss a fine structure of bifurcations for systems with various
potential energy surfaces. \cite{23,24,dos,6,Mamaev,Geldart} In
Section II, we introduce two-dimensional model potential energy
surfaces for a pair of interacting particles. In Section  III, we
study the effect of temperature on the tunneling rate. In Sections
IV and V, we calculate the rate for the parallel and antiparallel
tunneling and provide the analysis of the origin of bifurcation.
The effect of a promoting mode (an environment) is studied in
Section VI.

\section{\label{Sec2}Two-dimensional potential energy surfaces}

Consider two charges that tunnel in two independent double-well
potentials $U(q_1)$ and $U(q_2)$ presented as follows:
\cite{12,23,24,dos}
\begin{eqnarray}
\label{1}
\tilde{U}(q_i) = \frac{1}{2}{\omega^2(q_i+a)^2}\theta(-q_i)
\nonumber \\
 + \left[-\Delta I +
 \frac{1}{2}{\omega^2(q_i-b)^2}\right]\theta(q_i),
 \quad i=1,2,
\end{eqnarray}
where the sum $a+b$ determines the length of a "link" in the
corresponding macrocluster fragment; $\Delta I =
\omega^2(b^2-a^2)/2$ is a bias (an asymmetry parameter of the
potential); $\theta(q_i)$ is a step function, and $\omega$ is the
frequency (see discussion in  Ref.~\onlinecite{23}). The mass is
absorbed into the definition of $q$.

The interaction between two charges, e.g., protons, is considered
in a dipole-dipole approximation \cite{24}
\be\label{2}
V_{\mathrm{int}}(q_1,q_2) = - \frac{\alpha}{2}(q_1-q_2)^2,
\ee
where $\alpha$ is a positive constant. We use the same interaction
potential as in Ref.~\onlinecite{11}.

Thus, the total two-dimensional potential energy surface for
parallel tunneling normalized by $\omega^2$, is given by
\begin{eqnarray} U_{\mathrm{p}}(q_1,q_2) =
\frac{2\tilde{U}_{\mathrm{p}}(q_1,q_2)}{\omega^2} =
 (q_1+a)^2\theta(-q_1)
 \nonumber\\
 + \left[-(b^2-a^2)+(q_1-b)^2\right]\theta(q_1)
 + (q_2+a)^2\theta(-q_2)
 \nonumber\\
 + \left[-(b^2-a^2) + (q_2-b)^2\right]\theta(q_2)
 - \frac{\alpha^*}{2}(q_1-q_2)^2.
 \label{3}
\end{eqnarray}
Here, $\alpha^* = {2\alpha}/{\omega^2}$ is the dimensionless
parameter, $\alpha^* <1$, $\alpha \simeq e^2/(\varepsilon R_0^3)$,
$e$ is the electron charge, $R_0$ is the separation distance
between the reaction coordinates $q_1$ and $q_2$ of the tunneling
particles; $\varepsilon$ is the dielectric constant. The form of
potential energy surface (\ref{3}) is shown in Fig.~1.

\begin{figure}[tbp!]
\begin{center}
\includegraphics[width=0.5\textwidth]{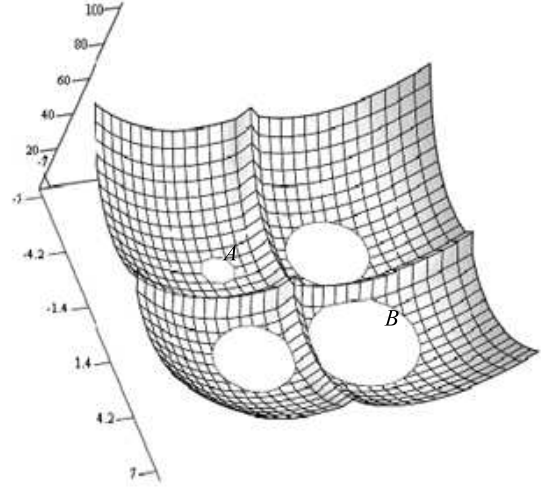}
\end{center}
\caption{ \label{Fig1} Asymmetric potential energy surface
(\ref{3}) for  parallel tunneling; $a=2$, $b=2.5$, $\alpha^*
=0.0001$. $A$ and $B$ indicate the initial and final states of the
particles, respectively. The minimum of the potential at $B$ is
lower than that of at $A$. The other two (intermediate) minima are
lower than those of at $A$ and higher than at $B$.}
\end{figure}

For antiparallel transfer, the two-dimensional potential energy
with the interaction term can be defined as
\begin{eqnarray} U_{\mathrm{a}}(q_1,q_2) =
\frac{2\tilde U_{\mathrm{a}}(q_1,q_2)}{\omega^2} =
 (q_1+a)^2\theta(-q_1)
 \nonumber\\
 + \left[-(b^2-a^2) + (q_1-b)^2\right]\theta(q_1) +
 (q_2-a)^2\theta(q_2)
  \nonumber\\
 + \left[-(b^2-a^2) + (q_2+b)^2\right]\theta(-q_2) -
 \frac{\tilde\alpha^*}{2}(q_1-q_2)^2,
 \label{4}
\end{eqnarray}
where  $\tilde\alpha^* = {2\tilde\alpha}/{\omega^2}$ is a
dimensionless parameter, $\tilde\alpha^* <1$. The potential
(\ref{4}) is depicted in Fig.~2. The main difference between the
potential energy surfaces (3) and (4) is in the initial location
of the second particle, $\pm a$.

\begin{figure}[tbp!]
\begin{center}
\includegraphics[width=0.5\textwidth]{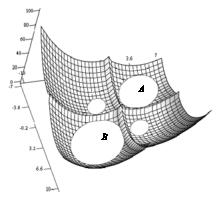}
\includegraphics[width=0.5\textwidth]{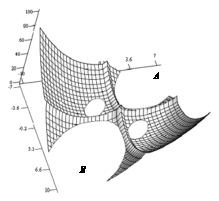}
\end{center}
\caption{ \label{Fig2} Asymmetric potential energy surface
(\ref{4}) for antiparallel tunneling; (a) $a=2$, $b=2.3$,
$\tilde{\alpha}^* =0.1$ (top panel); (b) $a=2$, $b=2.3$,
$\tilde{\alpha}^* =0.5$ (bottom panel). $A$ and $B$ indicate
initial and final states of the particles. The minimum of the
potential at $B$ is lower than that of at $A$. The other two
(intermediate) minima are higher than those of at $A$ and $B$.}
\end{figure}

The potential energy (\ref{3}) can be referred to as "parallel"
while the potential energy (\ref{4}) as "antiparallel".  We also
define a "symmetric" potential energy as a particular case of the
potential (\ref{4}) under the condition $a=b$, i.e.,
\begin{eqnarray}
U_{\mathrm{s}}(q_1,q_2) = \frac{2\tilde
U_{\mathrm{s}}(q_1,q_2)}{\omega^2} =
 (q_1+a)^2\theta(-q_1)
\nonumber\\
 + (q_1-a)^2\theta(q_1) +
 (q_2-a)^2\theta(q_2) + (q_2+a)^2\theta(-q_2)
 \nonumber\\
 - \frac{\tilde\alpha^{**}}{2}(q_1-q_2)^2,
 \label{5}
\end{eqnarray}
where the constant $\tilde\alpha^{**}<1$. The potential energy
(\ref{5}) is presented in Fig.~3.

\begin{figure}[tbp!]
\begin{center}
\includegraphics[width=0.5\textwidth]{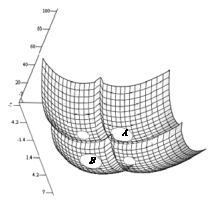}
\includegraphics[width=0.5\textwidth]{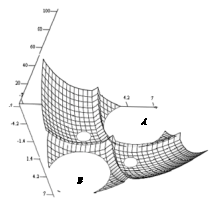}
\end{center}
\caption{ \label{Fig3} Symmetric potential energy surface
(\ref{5}); (a) $a=2$, $b=2$, $\tilde\alpha^{**} =0.1$ (top panel);
(b) $a=2$, $b=2$, $\tilde\alpha^{**} =0.5$ (bottom panel). $A$ and
$B$ denote the initial and final states of the particles. The
minimum of the potential energy at $B$ is equal to that of at $A$.
The other two (intermediate) minima are higher than those of at
$A$ and $B$.}
\end{figure}

As shown in Refs.~\onlinecite{24, dos, Mamaev}, such model
potential energy surfaces well describe the dynamics in
porphyrins.

In many practical cases, the effect of a bath on particles'
tunneling should be also included. In the next section we consider
two particles embedded into a harmonic medium and linearly
interacting with the bath modes.

\section{\label{Sec3} Two-particle transition probability}

We assume that two particles independently interact with a
harmonic bath. Such interaction is considered in a bilinear
approximation. The dynamics of the environment is described by the
oscillator Hamiltonian (we use $\hbar=1$, $k_B=1$ units  with the
oscillators'  masses  equal to one),
\be\label{6}
H_{\mathrm{ph}} = \frac{1}{2}\sum\limits_{i}\left(P_i^2 +
\omega_i^2Q_i^2\right).
\ee
Each of the tunneling particles (electrons or effective charges)
interacts with the oscillator bath in the following way:
\be\label{7}
 V_{\scriptsize\textrm{p-ph}}^{(1)}(q_1, Q_i) =
q_1\sum\limits_{i}C_iQ_i,\
 V_{\scriptsize\textrm{p-ph}}^{(2)}(q_2,
Q_i) = q_2\sum\limits_{i}C_iQ_i,
\ee
As in Ref.~\onlinecite{24}, we are interested in the transition
probability per unit time or, more precisely, only in its
exponential part which can be written in the Langer's form,
\be\label{8}
\Gamma = 2T\, \frac{\mathrm{Im}Z}{\mathrm{Re}Z}.
\ee
In such a consideration, metastable levels can be presented as
\be\label{9}
\Gamma = -2\, \mathrm{Im} E, \quad E = E_0 - i\Gamma/2.
\ee
It should be emphasized that Eq.~(8) is valid only for
temperatures below the crossover temperature while for higher
temperatures a different prefactor is required. \cite{Ankerhold}

Equation (\ref{8}) is obtained by generalizing the expression
(\ref{9}) to nonzero temperatures
\cite{7,8,9,10,11,12,13,14,15,16,17,18,19,20,21,22}
\be\label{10}
\Gamma
 =\frac{2\sum e^{-E_{0i}/T}\mathrm{Im}E_i}{e^{-E_{0i}/T}}
 =\frac{2T\, \mathrm{Im}\sum e^{-E_{i}/T}}
 {\mathrm{Re}\sum e^{-E_{i}/T}}
 = \frac{2T\, \mathrm{Im}Z}{\mathrm{Re} Z}.
\ee
Here, $i$ labels the energy levels in the metastable state, $Z$ is
the partition function of the system, and $T$ is the temperature.

To calculate $\Gamma$, it is convenient to present $Z$ in the form
of a path integral
\cite{7,8,9,10,11,12,13,14,15,16,17,18,19,20,21,22}
\be\label{11}
Z = \prod\limits_i \int\! Dq_1Dq_2DQ_i\exp[-S\{q_1,q_2,Q_i\}].
\ee
Here, $S$ denotes an underbarrier action for the total system. The
imaginary part $\mathrm{Im}Z$ is due to the decay of the energy
states in the initial well. The validity of this approximation
requires dissipation to be strong enough so that only an
incoherent decay occurs. \cite{Ankerhold} The imaginary part in
the partition function with a double well potential energy can be
also explained due to the strong dissipation to the bath. Indeed,
the particles do not come back to their initial state. Coherent
oscillations can happen only if the interaction with bosons is
weak enough \cite{13} or the bath is in nonequilibrium state.
\cite{d1,deb}

The integral (\ref{11}) can be performed over phonon coordinates
\cite{24}  resulting in
\begin{eqnarray}
S\{q_1,q_2\} = \int\limits_{-\beta/2}^{\beta/2} \!\!\! d\tau
\biggl[\frac{1}{2}\dot q_1^2 + \frac{1}{2}\dot q_2^2 + V(q_1,q_2)
 \nonumber\\
  + \int\limits_{-\beta/2}^{\beta/2} \!\!\! d\tau'
D(\tau-\tau')[q_1(\tau)+q_2(\tau)][q_1(\tau')+q_2(\tau')])
\biggr],
\label{12}
\end{eqnarray}
where
\be\label{13}
D(\tau) =
\frac{1}{\beta}\sum\limits_{n=-\infty}^{\infty}D(\nu_n)e^{i\nu_n
\tau},
\ee
$\beta = \hbar/(k_BT)$ is  an inverse temperature (below we assume
that $\hbar=1$ and $k_B=1$), $\nu_n = 2\pi n/\beta$ is the
Matsubara's frequency, and
\be\label{14}
D(\nu_n) = - \sum\limits_i\frac{C_i^2}{\omega_i^2 + \nu_n^2}.
\ee

A trajectory that minimizes the Euclidean action $S$ can be found
from the equations of motion. In particular, we embark on the
antiparallel tunneling
\begin{eqnarray}
-\ddot q_1 + \Omega_0^2q_1 + \tilde\alpha_1q_2 +
\!\!\!\!\int\limits_{-\beta/2}^{\beta/2}\!\!\!\!d\tau'
K(\tau-\tau')[q_1(\tau')+q_2(\tau')]
 \nonumber\\
 + \omega^2a\theta(-q_1)-\omega^2b\theta(q_1) = 0,\quad
\label{15}
\end{eqnarray}
\begin{eqnarray}
-\ddot q_2 + \Omega_0^2q_2 + \tilde\alpha_1q_2 +
\!\!\!\!\int\limits_{-\beta/2}^{\beta/2}\!\!\!\!d\tau'
K(\tau-\tau')[q_1(\tau')+q_2(\tau')]
 \nonumber\\
 - \omega^2a\theta(q_2)+\omega^2b\theta(-q_2) = 0. \quad
 \label{16}
\end{eqnarray}
In Eqs. (\ref{15}) and (\ref{16}) the kernel $K$ is defined by
\be\label{17}
K(\tau) =
\frac{1}{\beta}\sum\limits_{n=-\infty}^{\infty}\xi_ne^{i\nu_n\tau}.
\ee
Here, $\xi_n$ is the Green's function defined by Eq.~(14) without
the zero-frequency term,
\be\label{18}
D(\nu_n) = -\sum\limits_{i}\frac{C_i^2}{\omega_i^2} + \xi_n.
\ee
Thus, we seek for solutions of Eqs. (\ref{15}) and (\ref{16}) by
expanding the trajectories $q_1(t)$ and $q_2(t)$ into Fourier
series,
\be\label{19}
 q_1 =
\frac{1}{\beta}\sum\limits_{n=-\infty}^{\infty}q_n^{(1)}
e^{i\nu_n\tau},
\quad
 q_2 =
\frac{1}{\beta}\sum\limits_{n=-\infty}^{\infty}q_n^{(2)}
e^{i\nu_n\tau}.
\ee
Introducing the renormalized frequency and interaction constant,
\be\label{20}
\Omega_0^2 = \omega^2 - \sum\limits_{i}\frac{C_i^2}{\omega_i^2} -
\tilde\alpha,
 \quad
 \tilde\alpha_1 = \tilde\alpha -
 \sum\limits_{i}\frac{C_i^2}{\omega_i^2},
\ee
respectively, and substituting Eqs. (\ref{19}) into
Eqs.~(\ref{15}) and (\ref{16}), we obtain that for $n=0$,
\begin{eqnarray}
 q_0^{(1)} + q_0^{(2)} =
\frac{2\omega^2(a+b)\varepsilon}{\Omega_0^2+ \tilde\alpha_1},
\nonumber\\
 q_0^{(1)} - q_0^{(2)} =
-\frac{2\omega^2a\beta}{\Omega_0^2 - \tilde\alpha_1} +
\frac{4\omega^2(a+b)\tau_0}{\Omega_0^2 - \tilde\alpha_1},
\label{21}
\end{eqnarray}
and for $n\not=0$,
\begin{eqnarray}
 q_n^{(1)} + q_n^{(2)} =
\frac{2\omega^2(a+b)(\sin\nu_n\tau_1-\sin\nu_n\tau_2)}
{\nu_n(\nu_n^2
+\Omega_0^2 + \tilde\alpha_1 +2\xi_n)}, \nonumber\\
  q_n^{(1)} - q_n^{(2)} =
\frac{2\omega^2(a+b)(\sin\nu_n\tau_1+\sin\nu_n\tau_2)}
{\nu_n(\nu_n^2
+\Omega_0^2 - \tilde\alpha_1)}.
\label{22}
\end{eqnarray}
Here, we have introduced the following notation:
\be\label{23}
\varepsilon = \tau_1-\tau_2,
 \quad
 \tau_0 = (\tau_1+\tau_2)/2.
\ee
The time instants $\tau_1$ and $\tau_2$, at which the particles
pass the top points of the barrier, are determined from the
following equations:
\be\label{24}
q_1(\tau_1)=0,
 \quad
q_2(\tau_2)=0.
\ee
Equations (\ref{24}) allow us to change the argument of the
$\theta$-function. Namely, instead of the $q_1$- and
$q_2$-dependencies, we can use a time-dependent $\theta$-function.
This reduces Eqs.~(\ref{15}) and (\ref{16}) to a linear form.

Finally, substituting the trajectory determined from
Eqs.~(\ref{19}), (\ref{21}) and (\ref{22}) into Eq.~(\ref{12}), we
arrive at the following expression for the instanton action:
\begin{eqnarray}
S = \frac{4\omega^4a(a+b)\tau_0}{\Omega_0^2-\tilde\alpha_1}
 -
 \frac{\omega^4(a+b)^2\varepsilon^2}
 {\beta(\Omega_0^2+\tilde\alpha_1)}
 -
 \frac{4\omega^4(a+b)^2\tau_0^2}
 {\beta(\Omega_0^2-\tilde\alpha_1)}
\nonumber\\
 -
 \frac{8\omega^4(a+b)^2}{\beta}
 \sum\limits_{n=1}^{\infty}
 \left[
 \frac{\sin^2\nu_n\tau_0\cdot \cos^2(\nu_n\varepsilon/2)}
      {(\nu_n^2 + \Omega_0^2 - \tilde\alpha_1)\nu_n^2}\right.
 \nonumber\\
 \left.
 +\frac{\cos^2\nu_n\tau_0\cdot \sin^2(\nu_n\varepsilon/2)}
      {(\nu_n^2 + \Omega_0^2 + \tilde\alpha_1 + 2\xi_n)\nu_n^2}
 \right].\
 \label{25}
\end{eqnarray}

\section{\label{Sec4} Parallel particle tunneling}

For the case of parallel particle tunneling, the Euclidean action
$S$ can be determined similarly to antiparallel tunneling [see
Eqs. (\ref{15}) and (\ref{16})]. The trajectory minimizing the
Euclidean action (instanton), can be determined from the equations
of motion. As in the previous section, we seek for solutions of
these equations in the form of the Fourier expansion (\ref{19}).
The time instants $\tau_1$ and $\tau_2$ are determined by
Eqs.~(\ref{24}).

In the case of parallel tunneling particles [the potential energy
(\ref{3})], the resulting Euclidean action is given as follows:
\begin{eqnarray}
 S = 2a(a+b)(\tau_1+\tau_2)\omega^2
 -
 \frac{1}{\beta}\omega^2(a+b)^2(\tau_1+\tau_2)^2
 \nonumber\\
 -
 \frac{\omega^4(a+b)^2(\tau_1-\tau_2)^2}{(\omega^2-2\alpha)\beta}
 \nonumber\\
 -
 \frac{2\omega^4(a+b)^2}{\beta}
 \sum\limits_{n=1}^{\infty}
 \biggl[
 \frac{(\sin\nu_n\tau_1+\sin\nu_n\tau_2)^2}
 {\nu_n^2(\nu_n^2+\omega^2+\xi_n)}
  \nonumber\\
 +
 \frac{(\sin\nu_n\tau_1-\sin\nu_n\tau_2)^2}
 {\nu_n^2(\nu_n^2+\omega^2-2\alpha)}
 \biggr],
\label{26}
\end{eqnarray}
where $\xi_n$ is defined by Eq.~(\ref{18}).

Below, we use the following notation:
 $$\varepsilon = \varepsilon^*\omega=(\tau_1-\tau_2)\omega,$$
 $$\tau = 2\tau^*\omega =(\tau_1+\tau_2)\omega,$$
 $$\beta^* =\beta\omega/2,$$
 $$\alpha^*=2\alpha/\omega^2,$$
 $$b^*=b/a,$$
and assume that
 $$b \geq a.$$

In the absence of interaction with an oscillator bath, i.e., at
$\xi_n=0$, the action (\ref{26}) as a function of the parameters
$\varepsilon$ and $\tau$ yields
\begin{eqnarray}
S =
 \frac{(a+b)^2\omega}{2}
 \biggl\{
       \frac{4a\tau}{a+b}
       -\frac{\tau}{a+b}\left(1+\frac{1}{1-\alpha^*}\right)
\nonumber\\
 +
 \frac{(\tau\!\!-\!\!|\varepsilon|)\alpha^*}{1-\alpha^*}
 \!+\! \mathrm{coth}\beta^*
 \!-\! \mathrm{sinh}^{\!-\!1}\beta^*
    [
     \mathrm{cosh}(\beta^*\!\!-\!\!\tau)\mathrm{cosh}\varepsilon
\nonumber\\
+ \mathrm{cosh}(\beta^*\!\!-\!\!\tau)
     - \mathrm{cosh}(\beta^*-|\varepsilon|)
    ]
\nonumber\\
 -
 (1-\alpha^*)^{-3/2}
 \bigl[-
 \mathrm{coth}(\beta\sqrt{1-\alpha^*})
\nonumber\\
  + \mathrm{sinh}^{-1}(\beta\sqrt{1-\alpha^*})
    [
     \mathrm{cosh}((\beta^*\!\!-\!\!\tau)\sqrt{1-\alpha^*})
\nonumber\\
\times (\mathrm{cosh}(\varepsilon\sqrt{1-\alpha^*}) -1)
     +
     \mathrm{cosh}((\beta^*-|\varepsilon|)\sqrt{1-\alpha^*})
    ]
 \bigr]
\biggr\}. \quad
\label{27}
\end{eqnarray}
As soon as the trajectory is found, Eqs. (\ref{24}) can be
presented in the following form:
\begin{eqnarray}
\mathrm{sinh}\varepsilon [\,\mathrm{cosh}\tau \mathrm{coth}\beta^*
-\mathrm{sinh}\tau-\mathrm{coth}\beta^*] +
\nonumber \\
 \frac{1}{1\!-\!\alpha^*}\,
\mathrm{sinh}(\varepsilon\sqrt{1\!-\!\alpha^*})
[\mathrm{cosh}(\tau\sqrt{1\!-\!\alpha^*})\mathrm{coth}
(\beta^*\sqrt{1\!-\!\alpha^*})
\nonumber \\
-\mathrm{sinh}(\tau\sqrt{1-\alpha^*})
+ \mathrm{coth}(\beta^*\sqrt{1-\alpha^*})] =0, \nonumber\\
3-\frac{4}{1+b^*} - \frac{1}{1-\alpha^*} +
\mathrm{cosh}\varepsilon [\,\mathrm{sinh}\tau \mathrm{coth}\beta^*
-\mathrm{cosh}\tau
\nonumber \\
- 1] +\mathrm{sinh}\tau \mathrm{coth}\beta^* - \mathrm{cosh}\tau
\nonumber \\
+\frac{1}{1\!-\!\alpha^*}\,\mathrm{cosh}
(\varepsilon\sqrt{1\!-\!\alpha^*})
[\,\mathrm{sinh}(\tau\sqrt{1\!-\!\alpha^*})
\nonumber \\
\times
 \mathrm{coth}(\beta^*\sqrt{1\!-\!\alpha^*})
 - \mathrm{cosh}(\tau\sqrt{1-\alpha^*}) +1]
\nonumber \\
-\frac{1}{1-\alpha^*} [\,\mathrm{sinh}(\tau\sqrt{1-\alpha^*})\,
\mathrm{coth}(\beta^*\sqrt{1-\alpha^*})
\nonumber \\
-\mathrm{cosh}(\tau\sqrt{1-\alpha^*})]=0.\quad
\label{28}
\end{eqnarray}
Simple analytic solutions of Eqs. (\ref{28}) are obtained in the
particular case when
\begin{eqnarray}
\varepsilon = (\tau_1 - \tau_2)\omega =0, \quad \forall \beta,
\ \alpha <\omega^2/2, \nonumber \\
\tau_1 = \tau_2 = \frac{\tau}{2\omega} =
\frac{1}{2\omega}\,\mathrm{arcosh}\left[\frac{1-b^*}{1+b^*}\,
\mathrm{sinh}\frac{\beta\omega}{2}\right] +\frac{\beta}{4}.
\label{29}
\end{eqnarray}
However, a complete analysis requires numerical studies.

At sufficiently low temperatures, $\omega\beta \gg 1$, for
$1<b/a<3$, and
$$\frac{b-a}{2(b+a)} \leq
\frac{2\alpha}{\omega^2}< \alpha^*_c \equiv \frac{2(b-a)}{3b-a},
$$
we finally obtain, with the exponential accuracy,
\begin{eqnarray}
e^{-\tau\sqrt{1-\alpha^*}} \simeq [3 - \frac{4}{1+b^*} -
\frac{1}{1-\alpha^*}](1-\alpha^*)^{1/(1-\sqrt{1-\alpha^*})}
\nonumber \\
\times\left\{ 1+
(1-\alpha^*)^{1/(1-\sqrt{1-\alpha^*})}[-\frac{1}{1-\alpha^*}
+(3-\frac{4}{1+b^*} \right.
\nonumber \\
\left.
-\frac{1}{1-\alpha^*})/(1-\sqrt{1-\alpha^*})]
\right\}^{-1},
\nonumber \\
e^{-\varepsilon} \simeq [3 - \frac{4}{1+b^*} -
\frac{1}{1-\alpha^*}]e^{\tau\sqrt{1-\alpha^*}} +
\frac{1}{1-\alpha^*}. \quad
\label{30}
\end{eqnarray}

The solution (\ref{30}) is valid at
\be
\beta > \beta_c \equiv
\frac{\tau\sqrt{1-\alpha^*}}{\omega}.
\label{31}
\ee

We point out that an approximate solution can be found for large
values of the parameter $b^*$ (and small $\alpha^*$). However,
below we focus on the more important solution (\ref{30}). The
analysis indicates that there are no perturbative solutions of
Eqs.~(\ref{28}) at low temperatures and small $\varepsilon$.

For $\varepsilon =0$ [see Eq.~(\ref{29})], the action (\ref{27})
results in
\begin{eqnarray}
S_{\varepsilon=0} =
\omega(b^2-a^2)\mathrm{arcosh}\left[\frac{b^*-1}{b^*+1}\,
\mathrm{sinh}\frac{\omega\beta}{2}\right]
\nonumber \\
- \frac{1}{2}\omega^2(b^2-a^2)\beta + \omega(b+a)^2
\biggl[\mathrm{cosh}\frac{\omega\beta}{2}
\nonumber \\
-\left(1+\frac{(b^*-1)^2}{(b^*+1)^2}
\mathrm{sinh^2}\frac{\omega\beta}{2}\right)^{1/2}\biggr]
\mathrm{sinh}^{-1}\frac{\omega\beta}{2}.
\label{32}
\end{eqnarray}
The action (\ref{32}) coincides (up to a factor of two) with that
of calculated in Ref.~\onlinecite{23}. Thus, we have calculated
the two-particle Euclidean action for the case of synchronous
parallel motion of the two interacting particles.

In the symmetric case ($b^*=1$), the action (\ref{32}) reduces to
\be
S_{\!\! \scriptsize\begin{array}{c}\varepsilon\!=\!0\\[-2pt]
a\!=\!b
\end{array}}= 4\omega^2\mathrm{tanh}\frac{\omega\beta}{4}.
\label{33}
\ee
At $b^*>1$, the character of the temperature dependence is almost
the same.

At $\varepsilon\not=0$, the corresponding action $S$ can be
obtained by substituting Eq.~(\ref{30}) into Eq.~(\ref{27}) (for
brevity, this cumbersome expression is omitted). A simple analysis
shows that $S_{\varepsilon\not=0} < S_{\varepsilon=0}$. Moreover,
it appears that the difference $\Delta S = S_{\varepsilon\not=0} -
S_{\varepsilon=0}$ has a maximum at $\omega\beta \gg 1$.

\begin{figure}[tbp!]
\begin{center}
\includegraphics[width=0.4\textwidth]{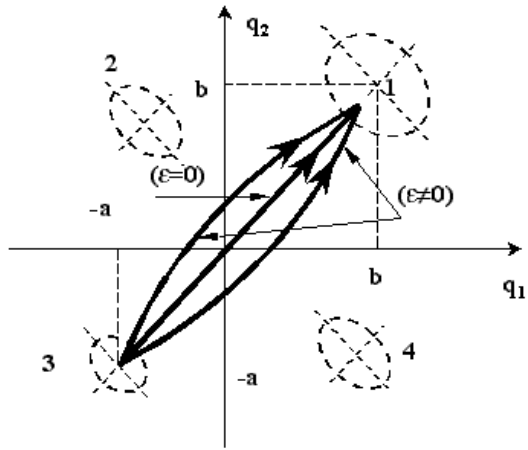}
\end{center}
\caption{\label{Fig4} Trajectories (a single path is characterized
by $\varepsilon=0$ and a splitted path is characterized by
$\varepsilon\not=0$) for two parallel moving particles, at
$\omega\beta \gg 1$; (1)-(4) label the projections of the minima
of potential energy (\ref{3}).}
\end{figure}

The tunneling paths (\ref{19}) are presented by the solutions
(\ref{29}) and (\ref{30}). At the critical point, $\beta=\beta_c$,
defined by Eq.~(\ref{31}), a relatively abrupt change in the
dynamics results in a splitting of the single trajectory
($\varepsilon=0$) into two  ($\varepsilon\not=0$), as shown in
Fig.~4.

\begin{figure}[tbp!]
\begin{center}
\includegraphics[width=0.45\textwidth]{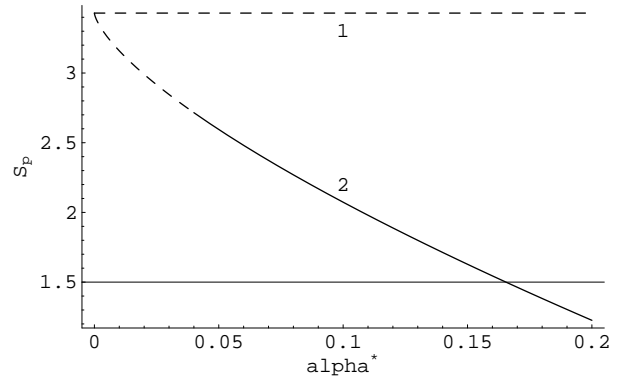}
\end{center}
\caption{ \label{Fig5} Instanton action as a function of the
interaction parameter $\alpha^*=2\alpha/\omega^2$ of the two
parallel moving particles, at $\omega\beta \gg 1 $; $S_{\mathrm
p}=S/(\omega a^2)$; (1) is the single trajectory; (2) is the split
trajectory.}
\end{figure}

At $\beta>\beta_c$, i.e., for temperatures below the critical
temperature, $T < T_c$, only a split trajectory
($\varepsilon\not=0$) occurs since $S_{\varepsilon\not=0} <
S_{\varepsilon=0}$ (see Fig.~5).

When $\beta<\beta_c$, i.e., at $T > T_c$, and $\alpha>\alpha_c$
[see Eq.~(\ref{30})], the solution is determined by a single
trajectory. We have also found that the single trajectory solution
holds in the whole temperature range for a symmetric potential
energy ($b^*=1$).

The above consideration is based on the analytic solutions
(\ref{29}) for $\tau_{1,2}$ near the critical point. A complete
analysis of Eq.~(\ref{28}) requires more detail.

\begin{figure}[tbp!]
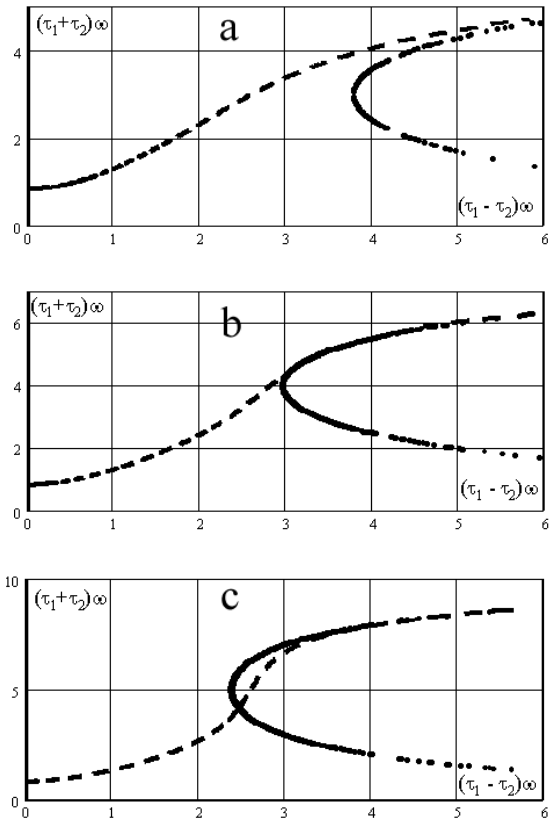

\begin{center}
\includegraphics[width=0.4\textwidth]{f6a}
 \vspace{14pt}

\includegraphics[width=0.4\textwidth]{f6b}
\vspace{14pt}

\includegraphics[width=0.42\textwidth]{f6c}
\end{center}
\caption{ \label{Fig6} Numerical solutions of transcendental
equations (\ref{28}). In addition to the studied solution
$\tau_1=\tau_2$, with larger $\beta$ there are additional
solutions, $\tau_1 \not= \tau_2$,  shown in panels (b) and (c).
Panel (b) reveals a {\em single} additional solution corresponding
to the lower value of the Euclidean action. Panel (c) shows {\rm
two} different additional solutions with the one corresponding to
the lower value of the action.}
\end{figure}

The numerical study of the set of transcendental equations
(\ref{28}) reveals remarkable features of two-dimensional
tunneling. The value of the critical parameter $\alpha_c$
decreases with temperature as shown in  Figs.~6, 7, and 8. The
temperature dependence of $\alpha_c$ indicates to the existence of
a finite low-temperature limit. It becomes clear that the weaker
the interaction between the tunneling particle, the lower
temperatures are required for the synchronous tunneling. Since the
synchronous tunneling is valid only for $T<T_c$ [see
Eq.~(\ref{31})] the dependence of $\beta_c$ on the interaction
parameter reveals that the minimum occurs at $\alpha^*=0.2$. The
nonmonotonic behavior of $\beta_c$ as a function of $\alpha^*$ can
be explained by the $\alpha^*$-dependence of $\tau$ as well.

\begin{figure}[tbp!]
\begin{center}
\includegraphics[width=0.45\textwidth]{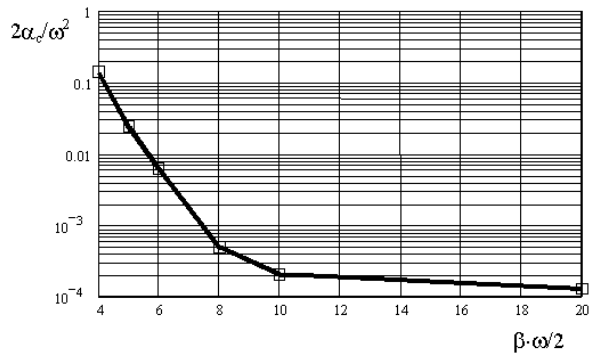}
\end{center}
\caption{ \label{Fig7} Parameter $\alpha_c$ as a function of
inverse temperature, at the fixed value of the frequency $\omega$
and asymmetry parameter $b^*$.}
\end{figure}

Additionally, the curve in Fig.~8 exhibits an anomalous behavior
(an increasing part of the curve) that can be explained in the
following way: a two-dimensional potential energy for parallel
tunneling is evidently deformed by an increase of the interaction
constant. Indeed, the minima of the potential surface become lower
and the distance between them is larger. Small deformations in the
potential energy surface change the contribution produced by the
temperature increase towards a synchronous character of the
tunneling. However, for sufficiently large deformations of the
potential, the situation is a quite opposite. The significant
deformations are in favor of synchronous transfer. Thus, an
increase in the interaction constant provides a similar effect
when temperature increases. Such a behavior takes place up to the
certain value of the interaction constant, beyond which the
potential energy surface becomes strongly perturbed. Thus, Figs.~7
and 8 are complimentary to  each other. Consequently, Fig.~8 can
be viewed as a bifurcation diagram. Indeed, the region below the
curve corresponds to the synchronous tunneling, while the region
above the curve corresponds to the asynchronous one.

\begin{figure}[tbp!]
\begin{center}
\includegraphics[width=0.45\textwidth]{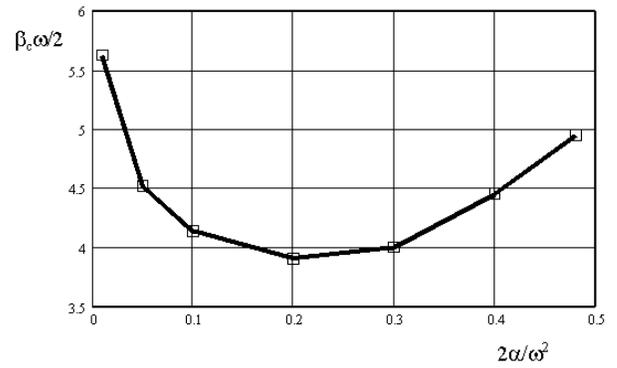}
\end{center}
\caption{ \label{Fig8} Parameter $\beta_c$ is presented as a
function of the interaction parameter of the two tunneling
particles.}
\end{figure}

Fig.~6 demonstrates how the critical parameter $\beta_c$ changes
from synchronous to asynchronous values with temperature. Figs.~6b
and 6c reveal the existence of the additional points, referred to
as the bifurcation points, corresponding to the change of
tunneling regimes. For the two bifurcation points shown in
Fig.~6c, only one  is physical because it corresponds to the lower
value of the action (the second point is metastable). However, for
sufficiently small differences between the values of the action at
these bifurcation points, {\em both} close points contribute, thus
leading to corresponding  "fluctuations" in the system during the
change of the regimes. For lower temperatures, such fluctuations
become negligible because two bifurcation points contributes
differently. Thus, a stable character of the asynchronous tunnel
transfer is achieved due to a much lower value of the action at
one of the bifurcation points.

For chemical reactions mentioned in Introduction, the effect of
the change in the tunneling regimes reveals as a {\em cleavage} in
the experimental temperature dependence of the tunneling constant.
Fine details of this instability have not been studied yet.
Nevertheless, the very existence of instability at the edge of the
bifurcation can be explained as the result of specific
fluctuations. A numerical analysis of the bifurcations in
antiparallel tunneling is given in the next section.

\section{\label{Sec5} Two-dimensional antiparallel tunneling}

For antiparallel tunneling of two particles [see the potential
energy (\ref{4})], the instanton action as a function of the
parameters $\varepsilon$ and $\tau$ is determined by
Eq.~(\ref{25}). For $\xi_n=0$, we obtain
\begin{eqnarray}
S = -\frac{\omega\tau(b^2-a^2)}{1-\tilde\alpha^*}
    -\frac{\omega(a+b)^2}{2}
    \left\{
    |\varepsilon|(1-\frac{1}{1-\tilde\alpha^*} )
    \right.
\nonumber \\
    +\frac{\mathrm{sinh}(|\varepsilon|\sqrt{1-\tilde\alpha^*})}
    {(1-\tilde\alpha^*)^{3/2}}
    - \mathrm{sinh}|\varepsilon|
    +\frac{\mathrm{cosh}(\varepsilon\sqrt{1-\tilde\alpha^*})+1}
    {(1-\tilde\alpha^*)^{3/2}}
\nonumber \\
     \times[\mathrm{sinh}(\beta^*\sqrt{1-\tilde\alpha^*})]^{-1}
     [\mathrm{cosh}((\beta^*-\tau)\sqrt{1-\tilde\alpha^*})
\nonumber \\
     -\mathrm{cosh}(\beta^*\sqrt{1-\tilde\alpha^*})]
\nonumber \\
         \left.
     +\frac{\mathrm{cosh}\varepsilon -1}{\mathrm{sinh}\beta^*}
     [\mathrm{cosh}(\beta^*-\tau) +\mathrm{cosh}\beta^*]
    \right\}.
\label{34}
\end{eqnarray}
The parameters $\varepsilon$ and $\tau$ are found from the
following set of equations [see Eq.~(\ref{24})]:
\begin{eqnarray}
 -\mathrm{sinh}\varepsilon[\mathrm{coth}\beta^*
 + \mathrm{cosh}\tau\mathrm{coth}\beta^* - \mathrm{sinh}\tau]
\nonumber \\
 + \frac{1}{1-\tilde\alpha^*}
 \mathrm{sinh}(\varepsilon\sqrt{1-\tilde\alpha^*})
   [
   \mathrm{coth}(\beta^*\sqrt{1-\tilde\alpha^*})
\nonumber \\
   -\mathrm{cosh}(\tau\sqrt{1-\tilde\alpha^*})
   \mathrm{coth}(\beta^*\sqrt{1-\tilde\alpha^*})
\nonumber \\
   +\mathrm{sinh}(\tau\sqrt{1-\tilde\alpha^*})
   ] = 0,
\nonumber \\
-1 - \frac{4}{(1+b^*)(1\!-\!\tilde\alpha^*)}
   + \frac{1}{1\!-\!\tilde\alpha^*}
\nonumber \\
   + (\mathrm{cosh}\varepsilon -1)
   (\mathrm{sinh}\tau\mathrm{coth}\beta^*
      -\mathrm{cosh}\tau)
\nonumber \\
   + \mathrm{cosh}\varepsilon
   + \frac{1}{1-\tilde\alpha^*}
     \left\{
     [\mathrm{cosh}(\varepsilon\sqrt{1-\tilde\alpha^*})+1]
     \right.
\nonumber \\
     \times[
     \mathrm{sinh}(\tau\sqrt{1-\tilde\alpha^*})
     \mathrm{coth}(\beta^*\sqrt{1-\tilde\alpha^*})
\nonumber \\
     - \mathrm{cosh}(\tau\sqrt{1-\tilde\alpha^*})
     ]
     \left.
     - \mathrm{cosh}(\varepsilon\sqrt{1-\tilde\alpha^*})
     \right\}
     =0. \quad
\label{35}
\end{eqnarray}
Simple analytic solutions of Eqs. (35) can be obtained in the
following form:
\begin{eqnarray}
    \varepsilon = (\tau_1-\tau_2)\omega =0, \quad \forall \beta, \
    \tilde\alpha < \omega^2/2,
\nonumber \\
    \tau_1 = \tau_2 = \frac{\tau}{2\omega}
\nonumber \\
    =
    \frac{1}{2\omega\sqrt{1\!-\!\tilde\alpha^*}}\
    \mathrm{arcosh}
    \left[
    \frac{1\!-\!b^*}{1\!+\!b^*}
    \mathrm{sinh}(\frac{\beta\omega}{2}\sqrt{1\!-\!\tilde\alpha^*})
    \right]
    + \frac{\beta}{4}.
\label{36}
\end{eqnarray}

Similarly to the case of parallel tunneling, we obtain that at low
temperatures, $\omega\beta \gg 1$, with the exponential accuracy,
\begin{eqnarray}
 e^{-\tau\sqrt{1-\tilde\alpha^*}} \simeq
   \frac{A(1-\tilde\alpha^*)^{1/\gamma}}
   {
   1- (1-\tilde\alpha^*)^{1/\gamma}({A}/{\gamma}
   - ({1-\tilde\alpha^*})^{-1})
   },
   \nonumber \\
   e^{\varepsilon} \simeq A e^{\tau\sqrt{1-\tilde{\alpha}^*}}
   - \frac{1}{1-\tilde\alpha^*}.
\label{37}
\end{eqnarray}
Here,
$$
A = -1 - \frac{4}{(1+b^*)(1-\tilde\alpha^*)}
       + \frac{3}{1-\tilde\alpha^*}, \quad
       \gamma = 1- \sqrt{1-\tilde\alpha^*},
$$
while $\tilde\alpha^*$, $b^*$, $\varepsilon$, and $\tau$ are
determined as for the parallel transfer.

The solution (\ref{37}) is valid at $\tilde{\alpha}^*_{c1} <
\tilde\alpha^* < \tilde{\alpha}^*_{c2}$ , where the lower and
upper bounds $\tilde{\alpha}^*_{c1}$ and $\tilde{\alpha}^*_{c2}$
are derived from a cumbersome transcendental equation (for brevity
it is not presented here). Particularly, in the symmetric case,
$b^*=1$, we obtain the condition in a simple analytic form, $1/4 <
2\tilde{\alpha}/\omega^2 < 1$.

Furthermore, an approximate solution can be found for large values
of the parameter $b^*=b/a$ (and small $\tilde{\alpha}^*$).
However, we restrict our analysis to the more important physical
solution (\ref{37}).

The $\beta$-dependent solution (\ref{37}) is valid for $\beta >
\tilde{\beta}_c$, where
\begin{eqnarray}
\tilde{\beta}_c \equiv - \frac{1}{\omega\sqrt{1-\tilde{\alpha}^*}}
        \ln
        \frac{A(1-\tilde{\alpha}^*)^{1/\gamma}}
        {1- (1-\tilde{\alpha}^*)^{1/\gamma}(\frac{A}{\gamma} -
        \frac{1}{1-\tilde{\alpha}^*})}.
\label{38}
\end{eqnarray}

At $\omega\beta \gg 1$, the solutions of Eqs.~(\ref{35}) can be
found perturbatively (for small $\varepsilon$) with given values
of the parameters $(b-a)/(b+a)$ and $\tilde\alpha^*$. At
$\varepsilon=0$ [the solution (\ref{36})], the action (\ref{34})
yields:
\begin{eqnarray}
S =
 \frac{\omega(b^2-a^2)}{(1-\tilde\alpha^*)^{3/2}}\,
 \mathrm{arcosh}
 \left[
 \frac{b-a}{b+a}\ \mathrm{sinh}
 \frac{\omega\beta\sqrt{1-\tilde\alpha^*}}{2}
  \right]
\nonumber \\
 - \frac{\omega^2\beta(b^2-a^2)}{2(1-\tilde\alpha^2)}
 + \frac{\omega(b+a)^2}{(1-\tilde\alpha^*)^{3/2}}
 \left[
 \mathrm{coth}\frac{\omega\beta\sqrt{1-\tilde\alpha^*}}{2}
 \right.
\nonumber \\
  \left.
 -\left(
   \mathrm{sinh}^{-2}\frac{\omega\beta\sqrt{1-\tilde\alpha^*}}{2}
   + \frac{(b-a)^2}{(b+a)^2}
 \right)^{\! 1/2}
 \right].
\label{39}
\end{eqnarray}

For the symmetric potential ($a=b$) and $\varepsilon=0$, we obtain that
(see Fig.~7)
\be
S = \frac{4\omega
 a^2}{(1-\tilde\alpha^*)^{3/2}}\
 \mathrm{tanh}\frac{\omega\beta\sqrt{1-\tilde\alpha^*}}{4}.
\label{40}
\ee

We do not present here a cumbersome expression for
$S_{\varepsilon\not=0}$ which one can obtain by substituting the
solutions $\tau$ and $\varepsilon$ into Eq.~(\ref{34}). A simple
analysis reveals that $S_{\varepsilon\not=0} > S_{\varepsilon=0}$
for $\beta > \tilde\beta_c$  and for relevant $\tilde\alpha^*$.
Similarly to parallel transfer, the tunnel paths can be found from
Eqs.~(\ref{36}) and (\ref{37}). These trajectories on the
$(q_1,q_2)$-plane are shown in Fig.~9.

\begin{figure}[tbp!]
\begin{center}
\includegraphics[width=0.4\textwidth]{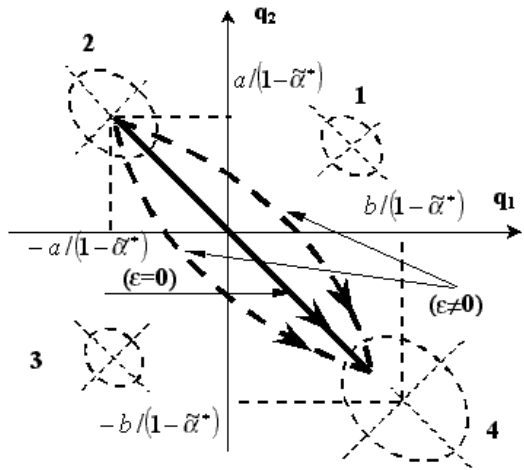}
\end{center}
\caption{ \label{Fig9} Trajectories (the basic path characterized
by $\varepsilon=0$ and the split one is characterized by
$\varepsilon\not=0$) at $\omega\beta \gg 1$ of two antiparallel
tunneling particles; (1)-(4) denote the projections of the minima
of the potential energy surface $U_{\mathrm{a}}(q_1,q_2)$ defined
by Eq.~(\ref{4}).}
\end{figure}

As for parallel tunneling and at $\beta>\tilde\beta_c$,  the
pair-tunneling changes from a single to double trajectory regime.
In contrast to the parallel tunneling, such a splitting occurs at
any values of the parameters of the potential. At
$\beta>\tilde\beta_c$, we have $S_{\varepsilon\not=0} >
S_{\varepsilon=0}$ and, therefore, $S_{\varepsilon=0}$ determines
the tunneling rate. At $\beta < \tilde\beta_c$, the two
degenerated trajectories are transformed into a single trajectory,
$q_1=-q_2$, corresponding to synchronous antiparallel transfer.

For single particle tunneling, there is only a single tunneling path
(instanton) minimizing the action. Hence, there are two
different types of trajectories for the pair of interacting
particles. Namely, the main contribution to the instanton action
is determined either by the single  or by the double-degenerated
paths depending on the value of $\beta$. We also point out that in
the case of the parallel tunneling, the particles do not
simultaneously pass the top points of the barrier,
$\tau_1\not=\tau_2$. This means that the tunneling transfer is
asynchronous.

At small values of the interaction parameter $\alpha^*$ [see
Eq.~(\ref{30})] and at temperatures such that $\beta<\beta_c$ [see
Eq.~(\ref{31})], there is no splitting of a single path
($q_1=q_2$). Therefore, the particles pass the top  of the
barriers  at the same instants ($\tau_1=\tau_2$). Consequently,
the transfer of the particles is synchronous. The temperature
dependence for the antiparallel transfer action is plotted in
Fig.~10 at various $\tilde\alpha^*$.

\begin{figure}[tbp!]
\vspace{14pt}
\begin{center}
\includegraphics[width=0.4\textwidth]{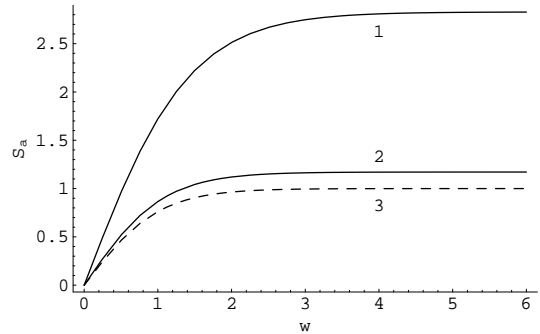}
\end{center}
\caption{ \label{Fig10} Instanton action for antiparallel
tunneling ($\varepsilon=0$, $a=b$) as a function of an inverse
temperature at the two different values of the interaction
parameter; $S_{\mathrm a}=S/(4\omega a^2)$, $w=\omega\beta/4$; (1)
$\tilde\alpha^*=0.5$; (2) $\tilde\alpha^*=0.1$; (3) a dashed line
corresponds to the action (\ref{33}) for the parallel transition
($\varepsilon=0$, $a=b$).}
\end{figure}

The type of the interaction given by Eqs.~(\ref{2})-(\ref{5}) is
such that it does not affect the motion along the "center of mass"
coordinate, $q_1=q_2$. For this reason, the Euclidean action is
independent of the interaction parameter as for parallel transfer.
Since the state of the interacting system characterized by a
maximal value of the relative coordinate, $q_1=-q_2$, is
preferable (as it provides the lower action), it becomes clear
that the instanton action decreases with the interaction parameter
in the parallel transfer along the degenerated tunnel
trajectories, and increases with the interaction parameter for the
antiparallel tunneling.

For antiparallel tunneling, synchronous transfer ($\tau_1=\tau_2$)
takes place, while asynchronous transfer is forbidden due to the
greater contribution to the Euclidean action (see Fig.~10).

The validity condition for weakly interacting
instanton-antiinstanton pairs, in the adiabatic approximation, was
discussed in Ref.~\onlinecite{23}.

\begin{figure}[tbp!]
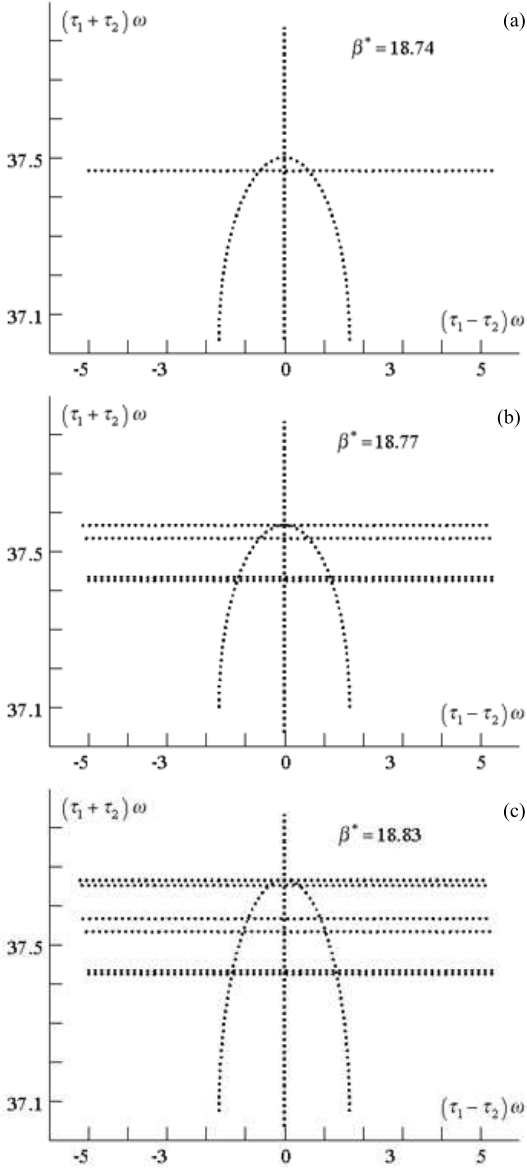

\vspace{14pt}
\begin{center}
\includegraphics[width=0.4\textwidth]{f11a}
\includegraphics[width=0.4\textwidth]{f11b}
\includegraphics[width=0.4\textwidth]{f11c}
\end{center}
\caption{ \label{Fig11} Numerical solutions of transcendental
equation (\ref{35}). In addition to the studied solution
$\tau_1=\tau_2$, with respect to $\beta$, there are the additional
solutions, $\tau_1\not=\tau_2$, shown in panels (a), (b) and (c),
that correspond to one, four, and six (pairs) additional
solutions, respectively.}
\end{figure}

The numerical analysis of transcendental equation~(\ref{35})
reveals interesting features for a transition region between the
tunneling regimes, i.e., a fine structure near the first
bifurcation point for the antiparallel transfer. The numerical
results are presented in Fig.~11. We found that, in addition to
the first bifurcation point characterized by the two solutions
(Fig.~11a), there exist additional bifurcation points at lower
temperatures, i.e., the four pairs  (Fig.~11b), the six pairs
(Fig.~11c), and there are even twelve pairs of additional
solutions at $\beta^*=19.2009$ $(\alpha^*=0.05)$, etc. We refer to
this phenomenon as multiplication of bifurcations or a {\em
cascade of bifurcations}. Such an effect resembles a scenario of
transition to chaos.

Although the synchronous regime is preferred due to the minimal
instanton action, in a certain temperature range its value is
comparable with those of corresponding to the cascade solutions.
As a result, {\em quantum fluctuations} of a nonregular character
occur in contrast to the parallel transfer. The antiparallel
tunneling is, thus, characterized by the instability of the
transition due to the  synchronous to asynchronous behavior. Such
instabilities are similar to continuous second-order phase
transition, while the parallel tunneling is viewed as a step
process similar to phase transition of a first order (see Fig.~6).
The dependencies shown in Figs.~7 and 8, $\beta_c(\alpha)$ and
$\alpha_c(\beta)$, for the antiparallel transfer are found to be
of the same character as those of the parallel transfer.

In summary, our investigation reveals a quite complicated fine
structure of the transition for parallel and antiparallel
tunneling of two particles with different degenerate trajectories
leading to the bifurcation cascade.

\section{\label{Sec6} Effect of a promoting mode}

In this section, we study the effect of a heat bath on tunneling
transition of two interacting particles. In many tunneling
reactions,  interaction with a vibrational subsystem can often be
approximated by interaction with a single vibrational mode (a
so-called promoting mode). As it follows from Eq.~(\ref{12}), a
heat bath affects only the dynamics of the center of mass
($q_1=q_2$). Therefore, in the case of an antiparallel motion, the
medium does not affect the rate constant, while for parallel
tunneling it essentially contributes to the transfer rate. For
both parallel and antiparallel transfer, bilinear interaction of
the particles with a single oscillator can make a qualitative
change in the character of tunneling.

At small values of the interaction parameter between the two
tunneling particles  dissipative effects become important.
\cite{23} In two dimensions the dissipative effects are more
pronounced for parallel rather than antiparallel tunneling. The
latter increases with  temperature with a considerable
contribution to the prefactor. In the present work, we are
interested in the tunneling rate assuming only exponential
evolution of the transition probability. However, nonexponential
evolution can occur in an nonequilibrium environment.
\cite{d1,d2,deb,dlw} Such a case is not discussed here.
Accordingly, a reservoir is assumed to be in thermodynamic
equilibrium, i.e., the tunneling transition is rather slow
compared to the thermodynamic relaxation.  Thus, we assume that
dissipation affects the value of instanton action only.

For the case of the antiparallel tunnel transfer, the action
(\ref{25}) can be calculated with the following vibronic Green's
function:
\be
D(\nu_n) = - \frac{C^2}{\nu_n^2 + \omega_L^2},
\label{41}
\ee
where $\omega_L$ is the frequency of the vibrational mode. After some
tedious calculations, one obtains the following expression for the
instanton action (\ref{25}):
\begin{eqnarray}
S = \frac{2\omega^4(a^2-b^2)\tau_0}{\Omega_0^2}
   - \frac{2\omega^4(a+b)^2}{\beta}
\nonumber \\
   \times\left\{
   \right.
   -\frac{\beta\mathrm{sinh}(\Omega_0(\beta/2-\tau_0))
         \mathrm{sinh}(\Omega_0\tau_0)}
         {2\Omega_0^3\mathrm{sinh}(\Omega_0\beta/2)}
\nonumber \\
   + \sum\limits_{i=1}^{2}\frac{\beta(\Omega_i^2-\omega_L^2)
   \mathrm{cosh}(\Omega_i(\beta/2-\tau_0))
                 \mathrm{cosh}(\Omega_i\tau_0)}
          {(-1)^i2\Omega_i^3(\Omega_1^2-\Omega_2^2)
          \mathrm{sinh}(\Omega_i\beta/2)}
\nonumber \\
   + \frac{\beta\varepsilon}{4(\omega^2 -2C^2/\omega_L^2)}
   - \frac{\beta\varepsilon}{4\Omega_0^2}
   + \frac{\beta\mathrm{sinh}(\varepsilon\Omega_0)}{4\Omega_0^3}
\nonumber \\
   +\sum\limits_{i=1}^{2}\frac{\beta(\Omega_i^2-\omega_L^2)
   \mathrm{sinh}(\varepsilon\Omega_i)}
         {(-1)^i4\Omega_i^3(\Omega_1^2-\Omega_2^2)}
   +\frac{\beta\mathrm{cosh}(\varepsilon\Omega_0)}
         {4\Omega_0^3\mathrm{sinh}(\beta\Omega_0/2)}
\nonumber \\
    \times [
            \mathrm{cosh}(\Omega_0(\beta/2-2\tau_0))
           - \mathrm{cosh}(\Omega_0\beta/2)
           ]
\nonumber \\
   -\sum\limits_{i=1}^{2}\frac{\beta(\Omega_i^2-\omega_L^2)
                              \mathrm{cosh}(\varepsilon\Omega_i)}
    {(-1)^i4\Omega_i^3(\Omega_1^2-\Omega_2^2)
    \mathrm{sinh}(\beta\Omega_i/2)}
\nonumber \\
    \times [
            \mathrm{cosh}(\Omega_i(\beta/2-2\tau_0))
            + \mathrm{cosh}(\Omega_i\beta/2)
           ]
  \left.
     \right\}.
\label{42}
\end{eqnarray}
Here, we have introduced the following notation:
\begin{eqnarray}
 \Omega_0^2 = \omega^2 -\alpha^2, \
 \Omega_1^2
 = \frac{1}{2}(\omega^2\!+\!\omega_L^2
   \!+\!\sqrt{(\omega^2\!-\!\omega_L^2)^2+8C^2}\ ),
\nonumber \\
 \Omega_2^2
 = \frac{1}{2}(\omega^2\!+\!\omega_L^2
   \!-\!\sqrt{(\omega^2\!-\!\omega_L^2)^2+8C^2}\ ).
\nonumber
\end{eqnarray}

For the parallel tunneling transition, a corresponding action can
be found in a similar way. The action (\ref{27}) as a function of
the parameters $\varepsilon^*=\tau_1-\tau_2$ and
$\tau^*=(\tau_1+\tau_2)/2$, with the vibronic frequency $\omega_L$
and coupling constant $C$, yields (see also Ref.~\onlinecite{23}):
\begin{eqnarray}
S \!=\!
 (a\!+\!b)(3a\!-\!b)\omega^2\tau^*\!
 -\frac{\omega^4(a\!+\!b)^2\varepsilon^*}{2(\omega^2\!-\!2\alpha)}
 -\frac{4\omega^2(a\!+\!b)^2\tau^{*2}}{\beta}
\nonumber \\
 -\frac{4\omega^4(a+b)^2\varepsilon^{*2}}{(\omega^2-2\alpha)\beta}
 -\frac{\omega^2(a+b)^2}{2\tilde\gamma}
 \sum\limits_{i=1}^{2}
 \frac{(-1)^i(\omega^2-x_{3-i})}{\sqrt{x_i}}
\nonumber \\
 \times
\left[
       \mathrm{coth}(\frac{\beta\sqrt{x_i}}{2})
       - \mathrm{sinh}^{-1}(\frac{\beta\sqrt{x_i}}{2})
\right.
         \left(
         \mathrm{cosh}[(\frac{\beta}{2}-2\tau^*)\sqrt{x_i}]
         \right.
\nonumber \\
          -\mathrm{cosh}[(\frac{\beta}{2}-\varepsilon^*)\sqrt{x_i}]
          +\frac{1}{2}\mathrm{cosh}[
          (\frac{\beta}{2}-2\tau^*-\varepsilon^*)\sqrt{x_i}]
\nonumber \\
\left.
         \left.
          +\frac{1}{2}\mathrm{cosh}[
          (\frac{\beta}{2}-2\tau^*+\varepsilon^*)\sqrt{x_i}]
         \right)
\right]
 + \frac{\omega^4(a\!+\!b)^2}{2(\omega^2\!-\!2\alpha)^{3/2}}
\nonumber \\
 \times
\left[
       -\mathrm{coth}(\frac{\beta}{2}\sqrt{\omega^2\!-\!2\alpha})
       +\mathrm{sinh}^{-1}(\frac{\beta}{2}\sqrt{\omega^2\!-\!2\alpha})
\right.
\nonumber \\
            \left(\!\!
           -\mathrm{cosh}[(\frac{\beta}{2}\!-\!2\tau^*)
           \sqrt{\omega^2\!-\!2\alpha}]
          +\mathrm{cosh}[(\frac{\beta}{2}\!-\!\varepsilon^*)
          \sqrt{\omega^2\!-\!2\alpha}]
            \right.
\nonumber \\
+\frac{1}{2}\mathrm{cosh}[
(\frac{\beta}{2}-2\tau^*-\varepsilon^*)\sqrt{\omega^2-2\alpha}]
\nonumber \\
\left.
              \left.
+\frac{1}{2}\mathrm{cosh}[
(\frac{\beta}{2}-2\tau^*+\varepsilon^*)\sqrt{\omega^2-2\alpha}]
              \right)
\right].\quad
\label{43}
\end{eqnarray}
Here, we have denoted
\begin{eqnarray}
x_{1,2} = \frac{1}{2}(\omega^2+\omega_L^2+\frac{C^2}{\omega_L^2})
          \mp \frac{1}{2}\tilde\gamma,
\nonumber \\
\tilde\gamma =
\sqrt{(\omega^2+\omega_L^2+\frac{C^2}{\omega_L^2})^2
          -4\omega^2\omega_L^2}.
\nonumber
\end{eqnarray}
For particular values of the interaction constant $\alpha$ and in
the absence of interaction with the oscillator bath, the critical
temperature $T_c$ (at which the synchronous and asynchronous
tunnel regimes interchange) is found from Eqs.~(\ref{31}) and
(\ref{38}). These equations can be generalized to non-zero
interaction with the promoting mode. Typically, the critical
temperature is found to be in the range from {\em 10 K} to {\em
400 K}. In glasses, $T_c$ can  be very small while for chemical
reactions it can be rather large. Additionally, $T_c$ depends on
the mean distance between the particles and, therefore, on their
concentration.

Quantum tunneling  is important \cite{24} when
$k_BT_c/(\hbar\omega) \leq 1$.  Therefore, the symmetry breaking
effects can take place at relatively high temperatures depending
on the ''frequency'' of a barrier. For example, for porphyrins the
critical temperature $T_c$ is estimated to be about {\em 200 K}.

\section{\label{Sec7} Conclusions}

In the single instanton approximation, we have calculated the
Euclidean action (\ref{12}) for the models characterized by the
different adiabatic potential energy surfaces, (\ref{3}),
(\ref{4}) and (\ref{5}), and made a detailed comparative analysis
of the tunneling rate for two interacting particles moving in
parallel or antiparallel within a dissipative environment. We have
also assumed exponential evolution of the transition probability.
\cite{23,d1,d2,deb,dlw}

We have shown that the change in a  tunneling regime from
synchronous to asynchronous transfer for parallel transition
occurs as a step process, similar to  phase transition of a first
order, while for the antiparallel transfer it resembles second
order phase transition.

We have explained the effect of a {\em cleavage} in the
experimentally observed \cite{3,4,5,Mamaev} temperature dependence
of the reaction rate for two tunneling particles. It has been
shown that the effect of symmetry breaking is stable for the
parallel and unstable for the antiparallel transfer, as it is
observed experimentally for some porphyrin systems.
\cite{3,4,5,Mamaev} We have found a complicated fine structure in
the bifurcation region due to the {\em quantum fluctuations} for
the parallel two-dimensional tunneling. For the antiparallel
tunneling, the contribution of four, six, twelve, etc., pairs of
trajectory becomes important resembling the transition to {\sl
chaos}.

Additionally, we have studied interaction of two particles with
phonons. Such coupling essentially modifies the antiparallel and
parallel transitions in different ways. As it follows from
Eq.~(\ref{12}), the interaction with the reservoir does not change
the dynamics of the center of mass for the antiparallel motion,
while it makes a significant contribution to the transfer rate for
the parallel transfer. Finally, Eq.~(\ref{31}) determines the
validity condition for temperatures beyond of which stable
two-dimensional synchronous tunneling
correlations of all the kinds occur.\\

\begin{acknowledgments}
The authors would like to thank A. I. ~Larkin and B. I. ~Ivlev for the
stimulating interest in this work.
\end{acknowledgments}

\end{document}